# COMMISSIONING AND FIRST RESULTS FROM THE FERMILAB CRYOMODULE TEST STAND*

E. R. Harms[†], M. Awida, C. Baffes, K. Carlson, S. Chandrasekaran, B. Chase, E. Cullerton, J. Edelen, J. Einstein-Curtis, C. Ginsburg, A. Grassellino, B. Hansen, J. Holzbauer, S. Kazakov, T. Khabiboulline, M. Kucera, J. Leibfritz, A. Lunin, D. McDowell, M. McGee, D. Nicklaus, D. Orris, J. Ozelis, J. Patrick, T. Petersen, Y. Pischalnikov, P. Prieto, O. Prokofiev, J. Reid, W. Schappert, D. Sergatskov, N. Solyak, R. Stanek, D. Sun, M. White, C. Worel, G. Wu, Fermilab, Batavia, Illinois, USA

*Abstract*

A new test stand dedicated to Superconducting Radiofrequency (SRF) cryomodule testing, CMTS1, has been commissioned and is now in operation at Fermilab. The first device to be cooled down and powered in this facility is the prototype 1.3 GHz cryomodule assembled at Fermilab for LCLS-II. We describe the demonstrated capabilities of CMTS1, report on steps taken during commissioning, provide an overview of first test results, and survey future plans.

## INTRODUCTION

LCLS-II is a next generation hard x-ray light source based on a superconducting RF electron linac operating in continuous wave regime. It is described at this conference in an invited talk [1]. As one of the partner labs Fermilab is responsible for the design of the 1.3 GHz Cryomodules (CM's) as well as assembly and testing for seventeen of the necessary thirty-five CM's. Additionally Fermilab is designing and will assemble and cold test two 8-cavity 3.9 GHz (third harmonic). Both the Cryomodule Test Facility and specifically the CMTS1 test stand have been described previously [2, 3].

## SUBSYSTEMS DESCRIPTION & STATUS

Guidance given by the LCLS-II project was to provide a test facility and carry out a testing program on a fast time scale. As such it was deemed best to design CMTS1 similarly to existing SRF test facilities at Fermilab such as the single cavity vertical test areas, HTS, and FAST where both single cavity and 8-cavity cryomodules have been tested and brought into routine operation. CMTS1 thus builds on evolving designs of the various subsystems and has features specifically to support continuous wave cryomodule operation.

Several of the necessary subsystems have been described previously including the cryogenics and distribution, interlocks, girders, and cave infrastructure and are not described further here.

### Vacuum

CMTS1 accommodates three separate vacuum systems: beamline, (warm) coupler, and insulating vacuum. The beamline vacuum system is an Ultra High Vacuum (UHV), low-particulate system that achieves characteristic pressures < $1 \times 10^{-8}$ Torr (warm) and < $1 \times 10^{-9}$ Torr (cold) via vacuum stations located at each end of the test stand. Pumpdown/vent-up of the vacuum stations is accomplished with a mass flow control vacuum cart to minimize the possibility of particle transport, and also includes an in-vacuum Faraday cup.

The coupler vacuum system is a conventional unbaked UHV system that achieves characteristic pressures <$1 \times 10^{-7}$ Torr (warm) and ~$1 \times 10^{-8}$ (cold, no RF), pumped by a single ion pump / titanium sublimation pump (TSP) unit.

The insulating vacuum system is characterized by large volume (~5000l) and high gas loads (up to $6 \times 10^{-2}$ Torr-l/s). Target vacuum levels are <$1 \times 10^{-4}$ Torr (warm) and < $1 \times 10^{-6}$ Torr (cold). A capacitance manometer is available for gas-independent pressure measurement and an RGA is available to differentiate between helium and air leaks. A stand-alone dry roughing system is used to pump between atmospheric pressure and $1 \times 10^{-2}$ Torr.

### Magnetic Flux

The pCM is instrumented with 13 Bartington™ *Mag-F* fluxgates [4, 5, 6]. Low field magnetic fields during cooldown are critical to maintaining high Q0. After transportation to and installation at CMTS, the pCM magnetic fields were no higher than 3.5 milliGauss (mG) compared to the specification of < 5mG. Although a demagnetization was not necessary at CMTS1, one was performed at 35 A [6] to confirm the absence of adverse effects of demagnetization on other systems at CMTS. Following this demagnetization, the field readings were all below 1.5 mG and all systems at CMTS were unaffected.

### Controls/Timing

General purpose controls for CMTS1 employs the Fermilab ACNET accelerator controls platform. The infrastructure for this system is mature, has a high level of support, and allows for relatively rapid deployment.

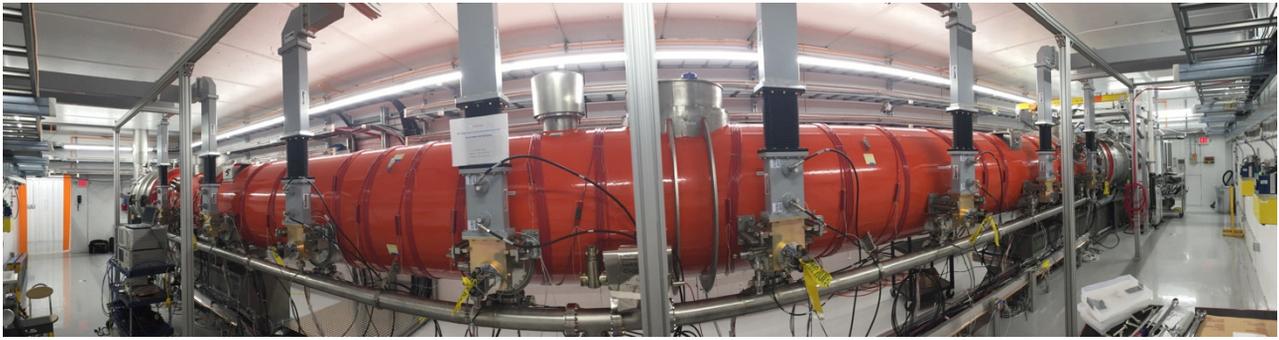

Figure 1: pCM installed in CMTS1 and readied for powering.

ACNET is flexible enough to allow interfaces to both specialized and commercial hardware and software.

CMTS1 uses a TCLK [7] based system for the generation and distribution of clock and timing signals. A block of 16 clock events is available for CMTS1 specific use. Synchronized timing gates for high level RF, RF interlocks, low level RF, and data acquisition are provided for each of the eight cavity systems.

*Interlocks*

The interlock system features eight independent subsystems, one per cavity/Solid State Amplifier (SSA) pair. Each system monitors several devices to detect fault conditions such as arcing in the waveguides or quenching of a cavity. These devices include field emission probes, photomultiplier tubes, infrared temperature sensors, and resistive temperature detectors. Each system can also detect fault conditions by monitoring the RF power seen at the cavity through a directional coupler. External permits that are common to all eight systems are also monitored. These are Cryo, Vacuum, Faraday cup window temperature, and Safety system permits. Lastly, an RF antenna monitors for stray non-ionizing radiation. In the event of a fault condition, each system is capable of removing RF signal to the SSA (via a fast RF switch) as well as turning off the SSA through an external permit line. Each interlock device signal is available for remote viewing and recording via a Fermilab designed digitizer board and MVME 5500 processor.

*Low Level RF*

The low-level RF system is comprised of four, 8 channel down-converters, four 2-channel up converters, local oscillator distribution, and control cards. These components translate the RF signals from 1.3GHz to 20 MHz and then to base-band for analysis and control calculations. The system can collect averaged data at 10 Hz and high-resolution data at 10 kHz.

The LLRF system can start up detuned cavities by using a frequency-tracking loop. The loop uses a PLL on the cavity probe signal to drive the RF signal to the resonant frequency of the cavity. If needed stepper motors and piezo tuners are then used to move the cavity resonant frequency to the master oscillator frequency. Once the cavity resonant frequency is equal to the master oscillator frequency, the LLRF system switches to driving the cavities at a fixed frequency and begin regulating the cavity amplitude and phase.

*High Level RF*

Each cavity is independently powered by its own 4 kW SSA. Amplifiers were delivered and verified functional in late 2015/early 2016. Directional couplers at three discrete locations between the SSA and Fundamental Power Coupler allow for several independent measures of the forward and reflected powers. Prior to being mated to the pCM each RF section was terminated in a calorimetric load and calibration measurements made.

*Instrumentation*

In total of order 460 channels of instrumentation are available on the pCM; 300 such channels are planned for the production CM's. In addition to that described elsewhere this instrumentation includes thermometry and pressure monitors and magnet current and resistive voltages used for magnet quench detection. Virtually all are interfaced to the controls system for monitoring, control, and data archival.

## INSTALLATION & FIRST RESULTS

In order to meet the strict performance requirements of LCSL-II, minimum acceptance criteria for both the prototype and production CM's have been established [8]. These largely drive the testing program. The primary criteria for the pCM include:

- Individual cavity gradient = 16 MV/m
- Field emission onset $\geq$ 14 MV/m
- Average cavity $Q_0 \geq 2.5 \times 10^{10}$
- Magnet coils powered to 20 A without quench
- Beamline vacuum at least $1 \times 10^{-10}$ Torr at 2K.

*Installation*

The pCM was delivered to the CMTF building on 20 July 2016. During its transport motion was monitored and no acceleration in excess of 0.2 g was detected, well below the self-imposed limit of 0.5 g [9].

Following several days of preliminary checks and preparation activities, the cryomodule was moved into the test cave by crane and vacuum and cryogenics connections made and checked. Figure 1 depicts the pCM once installed in CMTS1. Full checkout of all instrumentation and RF systems were also done in parallel as much as possible.

*Coupler Warm Conditioning*

Coupler warm conditioning was done at room temperature with isolated vacuum in the cryomodule and cavities off resonance by >100 kHz. In the first test the cavity #4 coupler was processed at 3kW power for about 20 hrs. RF processing continued for seven couplers in parallel a day later. The output power delivered to each coupler simultaneously varied from 1.8 to 3 kW owing to variations in amplifier gain. After ~16 hours of processing the power was increased up to a maximum of 4 kW for an additional six hours.

No multipactoring or breakdown were observed during these tests. Cavity vacuum was insensitive to the power level. The common warm coupler vacuum reached a maximum $3.X10^{-7}$ Torr and gradually improved during processing. Figure 2a provides summarized warm conditiong of Coupler #2 while 2b compares temperatures as a function of power.

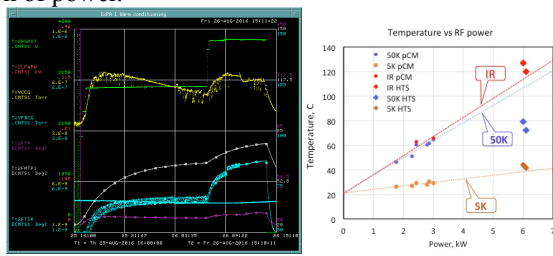

Figure 2a, b: Room Temperature processing of the 7 couplers in parallel. The left image (a) shows results for coupler #2 with the highest RF power (3 kW and then 4 kW): RF power (green), coupler vacuum (yellow) 5 K flange temperature (white), temperature of inner conductor of coupler (cyan). Image on the right (b) shows temperatures on different locations vs. RF power.

*Initial Cold Tests*

Cooldown from room temperature to 50 K was achieved over two days in a controlled fashion. 'Fast' cooldown to 4 K followed almost immediately thereafter and was accomplished in ~15 minutes with helium mass flow of 30 grams/sec. Complete cooldown to 2K has yet to be fully realized, but is expected by the time of this conference.

With the pCM stable at 2.1 K/30 Torr it has been possible to exercise all cavity tuning systems, verify their functionality, and bring cavities to resonance [10]. All cavities have had powers of sufficient amplitude applied to them to begin characterization. During single cavity testing #1 showed indications of HOM multipacting between 8 and 12 MV/m during single cavity testing accompanied by measurable radiation. When first energized at CMTS1 detectable radiation began at 400 Watts input power and continued to increase as shown in Figure 3a. Later, the forward power was set above the previous target power to 1300 Watts and the radiation levels significantly reduced as seen in Figure 3b.

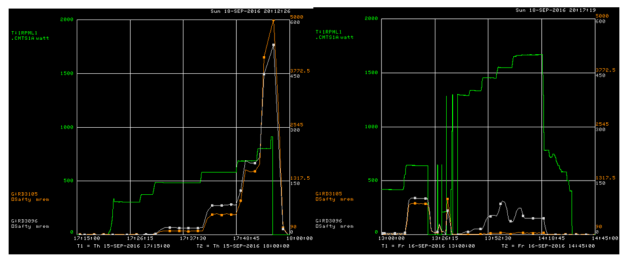

Figures 3a, b: Radiation levels vs. cavity input power for Cavity #1. In both plots the green trace is the output of the cavity SSA to a peak of 2 kW. Orange and Gray traces are radiation detectors situated near the cavity. Scales are the same in both plots with (b) indicating significant drop in radiation above 1.3 kW similarly to behaviour in previous tests.

Cavities #4 and #5 also exhibited field emission with applied power of order 1 kW. In neither case, however, did processing appear to occur. The remaining cavities have shown minimal field emission even at peak power.

*Next Steps*

Following this period of initial characterization, a multi-faceted extended test program to understand the performance of the pCM as fully as possible will be carried out. This is expected to last until late CY 2016. Production CM's will appear as early as December 2016. The 28-day production testing cycle is anticipated to commence in the first quarter of 2017 and continue into 2018.

## SUMMARY

CMTS1, Fermilab's newest cryomodule testing facility, has become operational and is now ready to support testing of CM's for LCLS-II. The prototype CM is installed, cold, and the first stages of powered cold testing are in progress. An aggressive LCLS-II CM testing program, 28 days testing cycle per CM, is anticipated.

## ACKNOWLEDGEMENTS

Assembly, commissioning, and now operation of CMTS1 reflects the dedication of a large cadre of talented people both at Fermilab and the institutions it has collaborated with over the years. The authors acknowledge the designers and builders of the cryomodules for which this test facility has been built, and especially the LCLS-II pCM at Fermilab.

Almost everyone involved in bringing CMTS1 to its current state is indebted to Helen Edwards and we honor her memory. Without her vision, technical acumen, and coaching, SRF technology would not be in the mature state found today at Fermilab.

## REFERENCES

[1] A. Burrill, "The LCLS-II SCRF Linac", presented at the 28[th] Linear Accelerator Conf. (Linac'16), East Lansing, MI, USA, paper MO2A01, this conference.

[2] E. R. Harms *et al.*, "Fermilab Cryomodule Test Stand Design & Plans". in *Proc. 17th Int. Conf. on RF Superconduc-

  *tivity (SRF2015)*, Whistler, Canada, September 2015, paper TUPB013, pp. 566-570.
[3] J.R. Leibfritz *et al.*, "Status and Plans for a Superconducting RF Accelerator Test Facility at Fermilab," in *Proc. Int. Particle Accelerator Conf. (IPAC'12)*, New Orleans, USA, May 2012, paper MOOAC02, pp. 58-60.
[4] S.K. Chandrasekaran, C. Ginsburg, 2016 Tesla Technology Collaboration meeting, Saclay, France.
[5] S.K. Chandrasekaran *et al.*, "Magnetic Field Management in LCLS-II 1.3 GHz Cryomodules", presented at *the 28th Linear Accelerator Conf. (Linac'16)*, East Lansing, MI, USA, September 2016, paper TUPLR027, this conference.
[6] S.K. Chandrasekaran *et al.*, "Demagnetization of a Fully Assembled LCLS-II Cryomodule", in preparation.
[7] David G. Beechy and Robert J. Ducar, "Time and Data Distribution Systems at the Fermilab Accelerator," *http://wwwbd.fnal.gov/controls/hardware_vogel/TCLK_Paper.pdf*
[8] J. Corlett *et al.*, "1.3 GHz Cryomodule Performance Requirements and Minimum Acceptance Criteria", SLAC, Menlo Park, CA, LCLSII-4.5-PP-0670-R0, July 2016.
[9] M. McGee, private communication, August 2015.
[10] W. Schappert *et al.*, "Performance of SRF Cavity Tuners at LCLS II Prototype Cryomodule at FNAL", presented at *the 28th Linear Accelerator Conf. (Linac'16)*, East Lansing, MI, USA, September 2016, paper THPRC017, his conference.